\documentclass[reprint,nofootinbib,citeautoscript, 
amsmath,amssymb, prx,floatfix, 
10pt]{revtex4-2}
\usepackage[ngerman,english]{babel}
\newif\ifAMStwofonts
\AMStwofontstrue

\def\gsim{~\rlap{$>$}{\lower 1.0ex\hbox{$\sim$}}}

\def\simpropto{\lower.2ex\hbox{$\; \buildrel \propto \over \sim \;$}}
\def\ltsim{\lower.5ex\hbox{$\; \buildrel < \over \sim \;$}}
\def\gtsim{\lower.5ex\hbox{$\; \buildrel > \over \sim \;$}}
\def\ltsim{\lower.5ex\hbox{$\; \buildrel < \over \sim \;$}}
\def\gtsim{\lower.5ex\hbox{$\; \buildrel > \over \sim \;$}}

\def\hunit{\,{\rm km\,s^{-1}\, Mpc^{-1}}}

\def\pmb#1{\setbox0=\hbox{#1}%
\kern-.025em\copy0\kern-\wd0
\kern.05em\copy0\kern-\wd0
\kern-.025em\raise.0433em\box0}

\def\simlt{\lower.5ex\hbox{$\; \buildrel < \over \sim \;$}}
\def\simgt{\lower.5ex\hbox{$\; \buildrel > \over \sim \;$}}

\newcommand{\beq}{\begin{equation}}

\newcommand{\eeq}{\end{equation}}
\def\beqa{\begin{eqnarray}}
\def\eeqa{\end{eqnarray}}
\def\fixit#1{}

\usepackage{graphicx}
\usepackage{dcolumn}
\usepackage{bm}
\usepackage{xcolor}
\usepackage{soul}
\usepackage[normalem]{ulem}
\usepackage[unicode=true,pdfusetitle,
 bookmarks=true,bookmarksnumbered=false,bookmarksopen=false,
 breaklinks=false,pdfborder={0 0 0},pdfborderstyle={},backref=false,colorlinks=true,citecolor=blue]
 {hyperref}
\newcommand{\HL}{Hubble--Lema\^itre }
\newcommand{\Lema}{Lema\^itre }

\begin{document}
\setstcolor{red}
\title{Superhorizon perturbations: A possible explanation of the Hubble--Lema\^itre Tension and the Large Scale Anisotropy of the Universe}
\author{Prabhakar Tiwari}
\email{ptiwari@nao.cas.cn}
\affiliation{National Astronomy Observatories, Chinese Academy of Science, Beijing  100101, P.R.China}

\author{Rahul Kothari}
\email{quantummechanicskothari@gmail.com}
\affiliation{Department of Physics \& Astronomy, University of the Western Cape, Cape Town 7535, South Africa}

\author{Pankaj Jain}
\email{pkjain@iitk.ac.in}
\affiliation{Department of Physics, Indian Institute of Technology, Kanpur-208016, India}
\newcommand{\red}[1]{\textcolor{red}{ #1}}
\date{\today}
\begin{abstract}
Current cosmological observations point to a serious discrepancy between the observed Hubble parameter obtained using direct  and cosmic microwave background radiation (CMBR) measurements. 
Besides this, the so called Hubble--Lema\^itre  tension, we also find considerable evidence in diverse cosmological observables that indicate violation of the cosmological principle. In this paper, we suggest that both these discrepancies are related and can be explained by invoking superhorizon perturbations in the Universe. We implement this by considering a single superhorizon mode and showing that it leads to both a dipole in large scale structures and a shift in the \HL parameter. Furthermore, the  shift is found to be independent of redshift up to a certain distance. This is nicely consistent with the data. 

\end{abstract}
\maketitle
\section{Introduction}
\label{sec:intro}
Around 90 years ago, Georges Henri Joseph \'Edouard Lema\^itre proposed that an expanding universe can explain the recession of nearby galaxies \cite{Lemaitre:1927}. With his expanding universe model, Lema\^itre derived the speed--distance relationship, the ``Hubble's Law"\footnote{Now called as the ``{\HL}  Law".}, and estimated the rate of cosmic expansion, i.e., the ``Hubble constant", equal to $645\hunit$ \citep{Peebles:1984Lemaitre,Way:2011,bergh:2011curious}.  He did this by combining  Gustaf Str\"omberg's redshift data  \citep{Stromberg:1925ApJ} (who relied mostly on Vesto Slipher’s work \citep{Slipher:1917}) and Hubble’s distances that were extracted using magnitudes \citep{Hubble:1926ApJ,Way:2011}. Soon after, Edwin Powell Hubble published his famous  paper \citep{Hubble:1929} where he and his assistant, Milton Humason used better stellar distance indicators such as Cepheid variables, novae and velocities. The velocity information was primarily extracted from the spectroscopic Doppler-shift observations due to Vesto Melvin Slipher \citep{Slipher:1917RadialVel}. This established a linear {relationship} between velocity \& distance and  determined the value for the cosmic expansion term, the Hubble's constant,  equal to $500\hunit$. Hubble’s remarkable observational relationship and Hubble constant value {were} obtained using only 24 nearby galaxies for which both measured velocities and distances were available with certain accuracy. Shortly after \Lema and Hubble’s discovery, cosmologists, including Einstein, became aware that far away objects are moving faster and thus the expanding universe model was established.

The theoretical and observational advances of cosmology have confirmed
a dark and exotic universe that is well described by the Friedman--\Lema--Robertson--Walker (FLRW) metric \citep{Friedmann:1922ZPhy,Friedmann:1924ZPhy,Lemaitre:1931,Lemaitre:1933ASSB,Robertson:1935,Robertson:1936II,Robertson:1936III,Walker:1937PLMS}, consisting of $\approx 70$\%  dark energy ($\Lambda$), $\approx 25$\% cold dark matter (CDM), and  with only $\approx 5$\% of familiar baryonic matter. Presently, the standard $\Lambda$CDM cosmological model that assumes zero spatial curvature {($\Omega_\mathrm{k}=0$) together with} isotropy and homogeneity, provides the simplest explanation of our universe. It provides a good fit to
 a large number of cosmological observations, such as the CMB radiation, primordial helium abundance, baryonic acoustic oscillations (BAO), galaxy clustering, Hubble parameter measurements etc. 
 
 Inspite of all these successes, there have been  several different observations showing significant tension with the standard $\Lambda$CDM. In particular, the `direct' measurements of the \HL parameter show a clear mismatch from the ones observed using `indirect' CMB measurements. Most notably, the recent direct measurement of \HL parameter from 
 Supernovae H0 for the Equation of State (SH0ES) collaboration, that uses Cepheid calibrated SNIa, yields $H_0=73.5\pm 1.4\hunit$ \citep{Reid:2019ApJ}. To the contrary, the Planck satellite using its precise CMB radiation fluctuation
 measurements \citep{Planck:2020_results2018} finds  $H_0=67.36\pm 0.54\hunit$. These two disagree with each other at $4.2 \sigma$ \citep{Anchordoqui:2019yzc} and this disagreement is widely known as \HL tension in the literature (see however \citep{Rameez:2019wdt} for a contrary view). For a review on the \HL tension, see \citep{Valentino:2021HL,Efstathiou:2020wxn,DiValentino:2020zio}. 
 
 There have been several simultaneous attempts  to calculate $H_0$ value using direct and indirect methods. On one hand, indirect methods usually employ CMB or Big Bang nucleosynthesis (BBN) with galaxy clustering measurements, \emph{viz.}, Sloan Digital Sky Survey (SDSS): Baryon Oscillation Spectroscopic Survey (BOSS), extended Baryon Oscillation Spectroscopic Survey (eBOSS) etc. They produce \HL parameter value roughly in agreement with the aforementioned Planck satellite value \citep{Pogosian:2020ApJ,Aiola:2020JCAP,Alam:2021}.  Whereas, the direct measurements, now extending across kpc to Gpc scales, include observations from Cepheids-SNIa \citep{Riess:2021}, Tip of the red-giant branch (TRGB)-SNIa \citep{Reid:2019ApJ}, Miras-SNIa \citep{Huang:2020}, Masers \citep{Pesce:2020ApJ}, Surface brightness fluctuations \citep{Blakeslee:2021ApJ}, Tully-Fisher relation \citep{Kourkchi:2020TF}, gravitational waves \citep{Gayathri:2020GW,Mukherjee:2020GW} etc., are roughly in agreement with SH0ES observation. The direct measurements also include observations of lensing systems --- for example the H0 Lenses in COSMOGRAIL's Wellspring (H0LiCOW) collaboration \citep{Wong:2020MNRAS}  finds $H_0=73.3^{+1.7}_{-1.8}\hunit$. Additionally, the Carnegie--Chicago Hubble Program, based on a calibration of  TRGB applied to SNIa \citep{Freedman:2019ApJ}, finds somewhat lower value of \HL parameter equal to $69.8\pm1.9 \hunit$.
Ever since \HL discovery, numerous extreme precision measurements of the \HL parameter have been carried out. Over the decades and to ever increasing distances, a variety of probes like SNIa standard candles  \citep{SupernovaSearchTeam:1998fmf,SupernovaSearchTeam:1998bnz,Perlmutter:1999,Kirshner:2004pnas,Betoule:2014}, improved stellar/Cepheid distance indicators \citep{Freedman:2001} etc., have been deployed to achieve this. These advancements have made the directly measured value of the Hubble parameter extremely accurate --- from Hubble's value of $H_0=500\hunit$ to {the present value} $H_0=73\pm \sim 2\hunit$. This has allowed us to make a close comparison between CMB derived $H_0$ value and the one from direct measurements.

To resolve this conflict between directly and indirectly measured $H_0$ values, there has been a flurry of proposals in the literature  (see \citep{Valentino:2021HL,Schoneberg:2021qvd} for a review of various solutions). These papers present various novel approaches such as the modification of the dark energy \citep{Bisnovatyi-Kogan:2020jmv, Ye:2020btb, Choi:2019jck, Panpanich:2019fxq, Karwal:2016vyq, Poulin:2018cxd, Berghaus:2019cls, Sakstein:2019fmf, Smith:2019ihp, Alexander:2019rsc}, introduction of non-standard neutrino interaction terms \citep{Blinov:2019gcj, Kreisch:2019yzn, Ghosh:2019tab, He:2020zns, Escudero:2019gvw,EscuderoAbenza:2020egd}, introduction of the fifth force \citep{Desmond:2019ygn}, emerging spatial curvature on account of the non-linear relativistic evolution \citep{Bolejko:2017fos}, modification of theory of gravity \citep{Vishwakarma:2020paa,Abadi:2020hbr,Shimon:2020mvl,Gurzadyan:2021jrw}, the modification of the $\Lambda$CDM by modifying or adding energy components to it \cite{Mortsell:2018mfj,Lin:2019qug, Kaya:2020hzn} etc.

In addition to the \HL tension, there exist other observations which also suggest a potential departure from the $\Lambda$CDM model. Some of these challenge the basic foundations of the model --- the cosmological principle. These include dipole anisotropy in radio polarization offset angles \cite{Jain:1998kf}, alignment of CMB quadrupole and octopole \cite{deOliveira-Costa:2003utu,Copi:2013jna,Aluri:2017mhj}, alignment of quasar polarizations \cite{Hutsemekers:1998}, dipole anisotropy in radio source counts \cite{Singal:2011,Gibelyou:2012,Rubart:2013, Tiwari:2014ni,Tiwari:2016adi,Colin:2017,Bengaly:2017slg} \& radio polarizations \cite{Tiwari:2015np}, bulk flow in X-ray clusters \cite{Kashlinsky:2010} etc. Remarkably, all these indicate a preferred direction close to the observed CMB dipole \cite{Ralston:2004}.  A recent study claims a dipole signal in quasar source counts at infrared frequencies which shows a deviation from the expected CMB dipole at $4.9 \sigma$ level \cite{Secrest_2021}. There also exist several claims of anisotropy in the Hubble constant \cite{Wiltshire_2013,Migkas:2021zdo,Biermann:1976,Luongo:2021nqh}. These observations suggest a potential departure from isotropy on the largest distance scales. A comprehensive discussion of such isotropy violations is given in \cite{Perivolaropoulos:2021jda}.

In this letter, we suggest that the two problems, i.e. (a) \HL tension and (b) the observed violation of isotropy at large distance scales are related. We propose a novel and elegant solution to both problems with a minimal modification of the $\Lambda$CDM model.

\section{A new proposal to relax \HL tension}
In the late-seventies, Grishchuk \& Zel'dovich \citep{Grishchuk:1978AZh,Grishchuk:1978SvA}  pointed out that long wavelength, i.e., superhorizon perturbations  to the metric could be significant without contradicting the observed temperature fluctuations of CMB. Such perturbations can explain the alignment of low-multipole moments of CMB \cite{Gordon:2005ai}. We also have constraints on  amplitudes and wavelengths of superhorizon perturbations from low-multipole moments of CMB \citep{Smoot:1991ApJ,Hinshaw:2003ApJS}.

To explain the implementation of superhorizon modes, we consider 
the conventional conformal  Newtonian gauge with the scalar perturbation to the flat FLRW metric given as,
\begin{equation}
\label{eq:metric}
\mathrm{d}s^2 =-(1+2\Psi)\mathrm{d}t^2 + a^2(t)(1-2\Phi) \delta_{ij}\mathrm{d}x^{i} \mathrm{d}x^{j},
\end{equation}
where $a(t)$ is the usual cosmological scale factor with $a_0=1$. The perturbation $\Psi$ to the temporal part of the metric  corresponds to the Newtonian potential. The scalar $\Phi$ is the perturbation to the spatial curvature. In the absence of anisotropic stress, $\Psi=\Phi$. A single adiabatic superhorizon mode perturbation, providing initial conditions for $\Psi$, in its simplest form can be modeled as \citep{Erickcek:2008a,Ghosh:2014,Das:2021JCAP}, 
\begin{equation}
\label{eq:SHmode}  
\Psi_\mathrm{p}= \alpha \sin(\kappa x_3+\omega),
\end{equation}
where {the subscript `p' is an  abbreviation for primordial,} $\alpha$ is the superhorizon mode amplitude, $x_3$ is the third component of the comoving position vector, $\kappa$ being the magnitude of the wave vector $\vec{k}$ and $\omega$ is a constant phase factor. Also, we have fixed  the coordinate such  that the wavevector $\vec{k}=\kappa {\hat{x}}_3$. This kind of simple superhorizon mode 
has been shown to significantly affect the large-scale distribution of matter and can potentially explain \citep{Ghosh:2014,Das:2021JCAP} the puzzling excess dipole signal observed in radio galaxy distribution \citep{Singal:2011,Gibelyou:2012,Rubart:2013, Tiwari:2014ni,Tiwari:2015np,Tiwari:2016adi,Colin:2017}
while simultaneously explaining the alignment of CMB quadrupole and octopole \cite{Gordon:2005ai}. The superhorizon mode in Eq. \eqref{eq:SHmode} 
introduces a perturbation to the gravitational potential between distant galaxies and us. This  effectively introduces corrections to observed redshifts of galaxies. Das et. al. \citep{Das:2021JCAP} show that a galaxy at redshift $z$ in the presence of a superhorizon perturbation $\Psi_\mathrm{p}$ will be observed at redshift $z_\mathrm{obs}$, such that \begin{equation}
\label{eq:zobs}
    1+z_\mathrm{obs} = (1+ z)(1+ z_{\rm Doppler})(1+ z_{\rm grav}),
\end{equation}
where $ z_{\rm Doppler}$ and $ z_{\rm grav}$ are respectively the redshifts
due to our velocity relative to large scale structure and 
potential perturbations introduced by superhorizon mode. The $z_\mathrm{obs}$ can be expanded and the leading monopole and dipole term can be written as, 
\begin{equation}
\label{eq:exp}
    z_\mathrm{obs}= \bar{z}+ \gamma \cos\theta+ ...,  
\end{equation}
Here $\theta$ is the polar angle of the spherical polar coordinate system,  $\gamma\equiv \gamma(z,\alpha,\kappa,\omega)$ and is small in comparison with $\bar{z}$, the monopole term. 
The monopole term $\bar{z}$,  given in \citep{Das:2021JCAP}, is
\begin{equation}
\label{eq:monopole}
    z_{\rm obs} \approx \bar{z} = z+ (1+z) [g(z)-g(0)] \alpha \sin \omega,
\end{equation}
{where $g(z)$ represents the redshift evolution of $\Psi(z)$ such that $\Psi(z)=g(z)\Psi_{\rm p}$.}

Ghosh \citep{Ghosh:2014} and Das et. al. \citep{Das:2021JCAP} successfully explain excess dipole signal observed with high redshift galaxies \citep{Singal:2011,Gibelyou:2012,Rubart:2013, Tiwari:2014ni,Tiwari:2015np,Tiwari:2016adi,Colin:2017} using superhorizon perturbations. They find that a superhorizon mode, for a range of $\alpha, \kappa$ values,  can consistently explain both the CMB and NVSS observations while remaining in harmony with others,  
\emph{viz.}, the dipole anisotropy in local \HL parameter measurements and local bulk flow observations. Das et al. \citep{Das:2021JCAP} note  that the phase $\omega=\pi$  conventionally maximizes the dipole signal in radio galaxies. They also provide sufficient details regarding  superhorizon perturbations and their applications to galaxy clustering, \HL parameter anisotropy measurements, bulk flow etc. In addition to this, they find a monopole contribution which is non-zero as long as the phase $\omega\ne\pi$. The choice $\omega=\pi$ is rather special and implies that we have a preferred position in the universe. There is no physical motivation for this and here we explore the implications of the monopole term. 
Remarkably, we find that the monopole term in Eq. \eqref{eq:monopole} potentially solves the \HL tension. We consider general values of phase $\omega$ and obtain redshift monopole term in Eq. \eqref{eq:monopole}. The \HL law (for nearby galaxies) is written as,
\begin{equation}
\label{eq:HL_law}
    {V_{\rm obs}}=H_0^{\rm obs} d, 
\end{equation}
where $V_{\rm obs}$ is the observed radial velocity of a galaxy at the proper distance $d$ and $H^{\rm obs}_0$ is the observed \HL parameter, corresponding to 
the observed redshift $z_\mathrm{obs}$. Thus we have (assuming ${z}_\mathrm{obs}\ll 1$)
\begin{equation}
    {z_{\rm obs}}= \frac{V_{\rm obs }}{c}=\frac{d}{c {(H_0^{\rm obs })}^{-1}}.\label{eq:zBarHubble}
\end{equation}
Analogously, in the equation for  redshift $z$, $V_{\rm obs}$ and $H^{\rm obs}_0$ would  respectively be replaced by $V$ and $H_0$. Here $H_0$ is the \HL parameter predicted from CMB measurements.

We note that the velocity measurements are carried out using Doppler shift and thus the apparent change in $z$ results a change in velocity. This eventually leads to a change in distance-velocity relation slope, i.e., \HL parameter value. 
Therefore, if the observed redshifts  $z_\mathrm{obs}$  
differ $z$ by 10\%, the directly measured value $H^{\rm obs}_0$ will also differ  from $H_0$ by the same amount. As the \HL tension is all about $\sim10$\% excess in directly measured \HL parameter value, we seek this change in the redshift dipole term given in Eq. \eqref{eq:monopole}. We find that a superhorizon mode with a range of $\alpha,\ \kappa$ and $\omega$ values can consistently explain the excess galaxy dipole and \HL tension. The set of values of these parameters is consistent with CMB, local bulk flow and local redshifts anisotropy limit.
 
\subsection{Possible values of parameters}
\begin{figure}
    \includegraphics[scale=0.5]{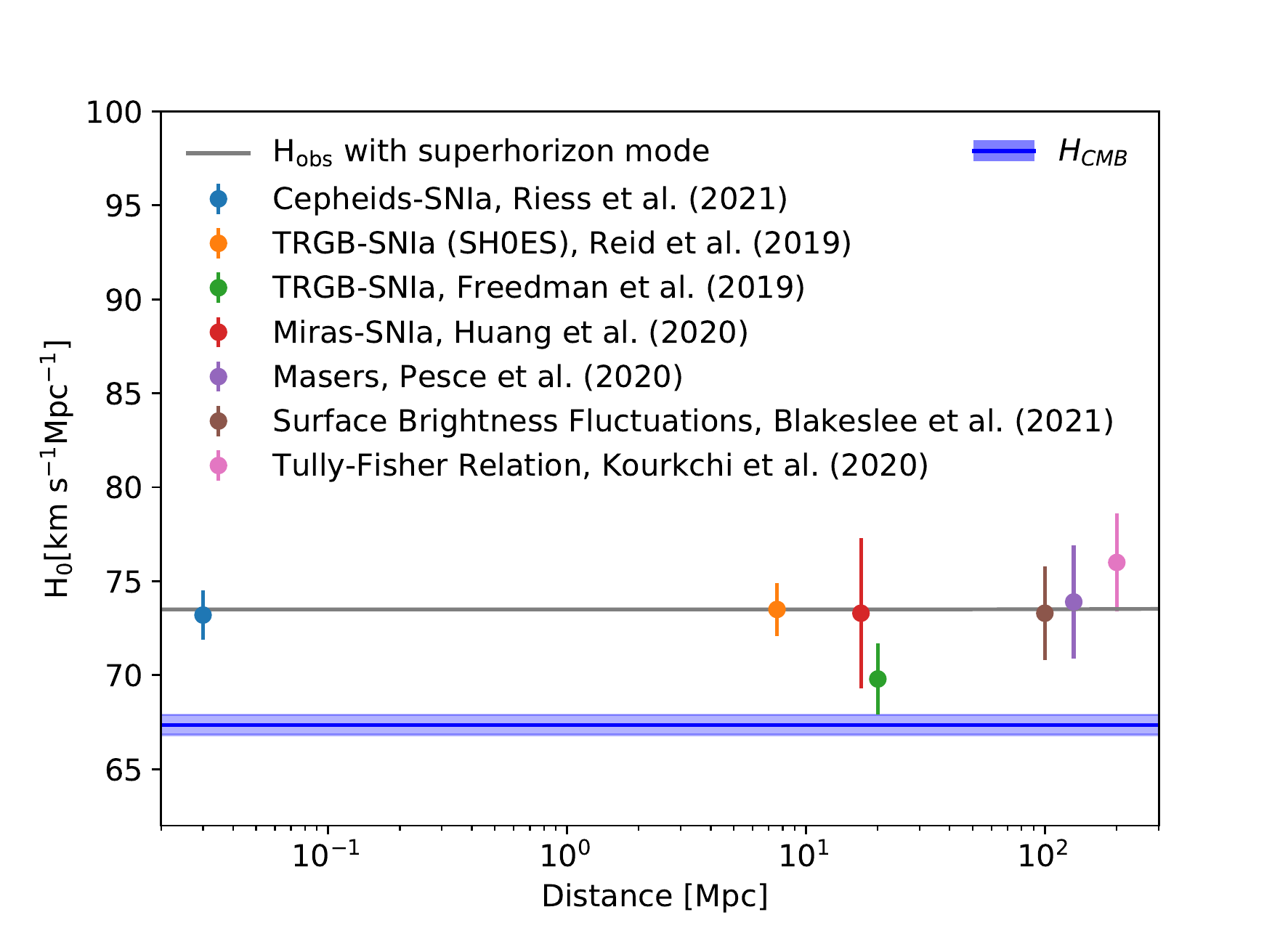}
    \caption{Apparent value of \HL parameter with direct measurements at different distances. It is noted that the directly measured \HL parameter remains roughly the same from distance as small as  kpc to hundreds of Mpc. The \HL measurements (data points) are obtained by combing many objects, the distance (x-axis) shown in figure corresponds to far most object used for analysis. All data points except CMB are direct measurements of \HL parameter.}
    \label{fig:H_dist}
\end{figure}

We consider Reid et al. \citep{Reid:2019ApJ} work and produce the \HL  measurements considering superhorizon perturbation mode. The details of the fitting procedure are given in the Appendix. The results are shown in Figure \ref{fig:H_Reid_etal}. Some  possible sets of values of  $\alpha, \kappa$ and $\omega$, that simultaneously explain the dipole from
NRAO VLA Sky Survey (NVSS) data\footnote{We have assumed dipole amplitude equal to value $0.0151$ \citep{Tiwari:2014ni}, and we have considered the number density, $N(z)\propto z^{0.74} \exp \left[ - \left(\frac{z}{0.71}\right)^{1.06}\right ]$, and galaxy bias $b(z)=0.33 z^2 + 0.85 z +1.6$ \citep{Adi:2015nb,Tiwari:2016adi}. We impose an upper redshift cutoff in the abundance of sources at z = 3.5. Note that this redshift cutoff in reference \citep{Das:2021JCAP} is z=2.}  and the \HL measurements are listed in Table \ref{tab:SH_par}. In general, a superhorizon mode with a phase $\omega$, with a `reasonable' amplitude $\alpha$ and wavelength $2\pi/\kappa$ can explain excess NVSS dipole and \HL tension. Furthermore, it is also noted that for a given superhorizon mode, the apparent \HL parameter roughly remains  the same for a very wide range of distances. The variation of {$H^{\rm obs}_0$} with distance is given in Figure \ref{fig:H_dist}. We add that superhorizon modes with CMB limits along with the NVSS dipole solution curve are consistent with local Hubble dipole and bulk flow observations \citep{Das:2021JCAP}. 

\begin{figure}
    \centering
    \includegraphics[scale=0.5]{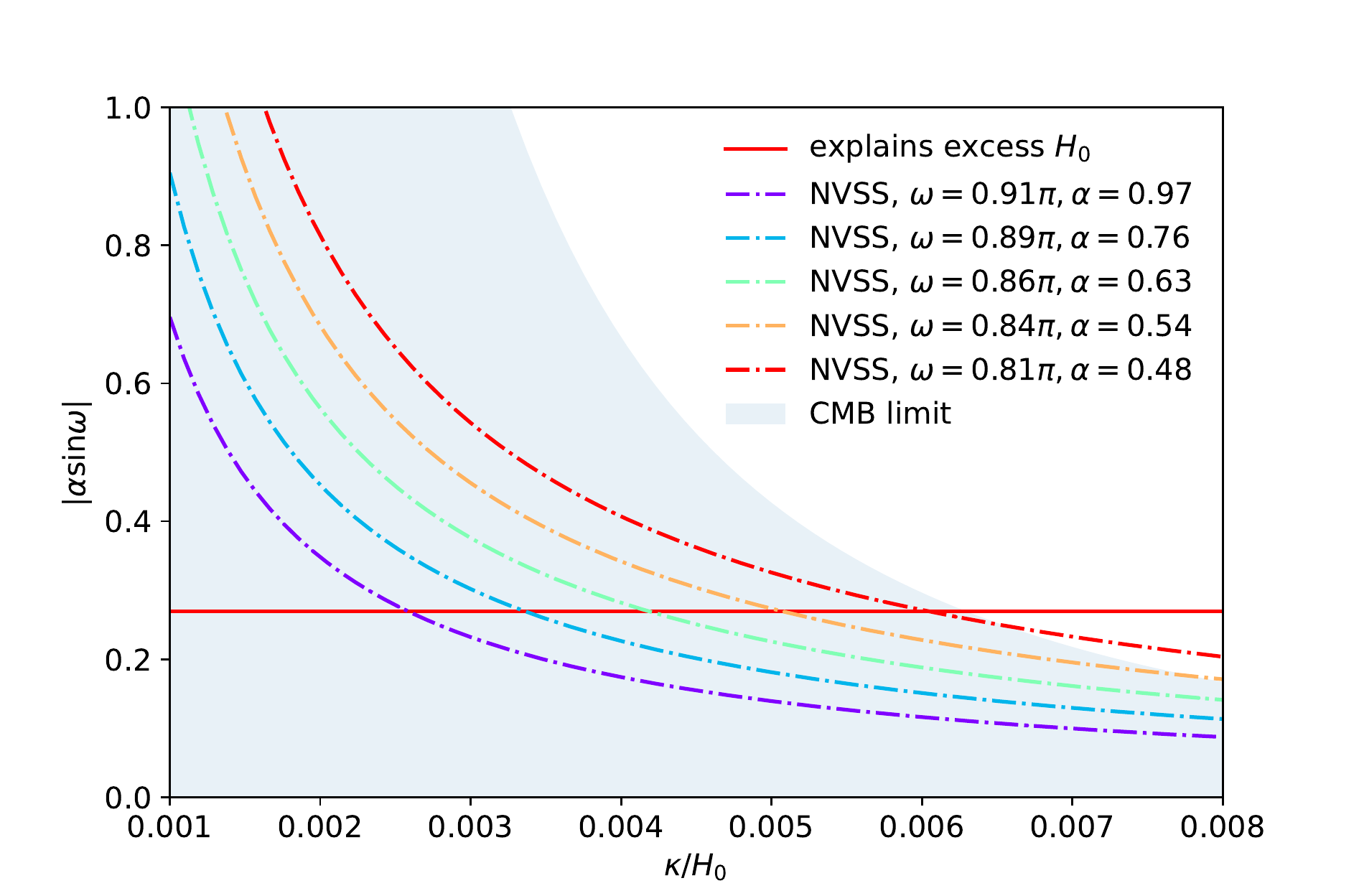}
    \caption{{Plot of $|\alpha \sin \omega |$ versus $\kappa/H_0$ for the observed \HL tension and NVSS dipole amplitude for various values of $\omega$. We have assumed dipole amplitude equal to $0.0151$ \citep{Tiwari:2014ni}. The horizontal line satisfies Reid et. al. \citep{Reid:2019ApJ} results, i.e.,  ${H^{\rm obs}_0}=73.5 \hunit$ at 7.57 Mpc, and the dotted dashed curves represent the NVSS excess dipole solution in $\kappa$, $\alpha$ plane with different values of phase $\omega$. The blue-shaded part denotes the region of parameter space that satisfies the CMB quadrupole constraint \citep{Erickcek:2008a}. Note that for $\kappa$ range we have explored, and phase, $\omega\ne \pi $, only the quadrupole constraints turn out to be relevant.}}
    \label{fig:H_Reid_etal}
\end{figure}

\begin{table}
      \begin{tabular}{| c | >{\hspace{1pc}} c <{\hspace{1pc}} | >{\hspace{1pc}} c <{\hspace{1pc}} |>{\hspace{1pc}} c <{\hspace{1pc}}|}
    \hline 
        &&&\\
         Superhorizon & $\omega$  &  $\alpha$ &  $\kappa \over H_0$ \\
          mode        &          &          &             \\
         \hline
         1. & $0.91\pi$ & 0.97 & $2.581\times 10^{-3}$ \hfill\\
         2. & $0.89\pi$ & 0.76 & $3.357\times 10^{-3}$ \hfill \\
         3. & $0.86\pi$ & 0.63 & $4.180\times 10^{-3}$ \hfill \\
         4. & $0.84\pi$ & 0.54 & $5.067\times 10^{-3}$ \hfill \\
         5. & $0.81\pi$ & 0.48 & $6.037\times 10^{-3}$ \hfill \\
         \hline
    \end{tabular}
    \caption{Some possible superhorizon modes $\Psi_\mathrm{p}$ with appropriate parameters  satisfying Reid et. al. \citep{Reid:2019ApJ} results, i.e.  ${H^{\rm obs}_0}=73.5 \hunit$ at 7.57 Mpc, and NVSS excess dipole simultaneously.}
    \label{tab:SH_par}
\end{table}

\section{Conclusion and Outlook}
Current cosmological observations suggest two discrepancies with the $\Lambda$CDM model. These are the \HL tension and the large scale anisotropy of the Universe. With more observations, these discrepancies have only grown stronger. In this paper, we have shown that both of these can be explained within the framework of a phenomenological model that assumes the existence of a superhorizon mode in the Universe. Such a mode is consistent with all existing cosmological observations and is already known to explain the CMB quadrupole-octopole alignment and the excess dipole in large scale structures. It is therefore quite fascinating that it can also explain the Hubble-Lemaitre tension, which \textit{apriori} appears to be a completely independent phenomenon. The model introduces only three new parameters, namely the amplitude, wavelength and the phase of the mode. Besides this, no special fine tuning of parameters is required for fitting the observables. It leads to a Hubble parameter which is approximately constant up to distances of order few hundred Mpc, which is nicely consistent with observations. 
The model is likely to make a wide range of cosmological predictions which can be tested in future. In particular, it will induce a small anisotropy along with a correlated isotropic shift in several other cosmological observables, such as BAO and the epoch of reionization, besides Hubble constant.  At large redshifts, the model predicts an interesting redshift dependence, which can also be tested in future studies \cite{Das:2021JCAP}, {this unique feature can potentially provide a strong evidence for the existence of superhorizon modes}. 

Theoretically, such a superhorizon mode may arise as a stochastic phenomenon, called spontaneous breakdown of isotropy \cite{Gordon:2005ai}. In this case it would be consistent with the cosmological principle. Another interesting possibility is that it may arise from an early phase of inflation. Inflation provides 
 the only known theoretical explanation for the observed isotropy and homogeneity of the Universe. Since we do not so far know how the Universe originated, it is natural to assume that at early time it may be described by an unknown inhomogeneous and anisotropic metric. During inflation, and essentially independent of the initial conditions, the metric becomes almost identical to the standard FLRW metric, possibly within one $e$-fold \citep{Wald:1983}. During such an early phase, when the Universe had not yet acquired its cherished properties of isotropy and homogeneity, it can generate modes which do not obey the cosmological principle \citep{Rath:2013}. Furthermore, there exist a wide range of parameters for which these modes can affect observations today \citep{Aluri:2012Pre-Inf}. Hence these observations, which appear to show deviations from $\Lambda$CDM, might offer a glimpse into a so far obscure early phase of the Universe and may be consistent with the Big Bang paradigm \cite{Rath:2013}.

\section*{Acknowledgements}
PT acknowledges the support by the National Key Basic Research and Development Program of China (No. 2018YFA0404503) and NSFC Grants 11925303 and 11720101004, and a grant of CAS Interdisciplinary Innovation Team. RK is supported by the South African Radio Astronomy Observatory and the National Research Foundation (Grant No.  75415). 

\appendix*
\section{Procedure}
\label{sc:methods}
In this Appendix we provide the details of the procedure used to fit the Hubble constant and the NVSS dipole.
Using Eq.  \eqref{eq:monopole}, we get 
\begin{equation}
\label{eq:asinw}
\alpha \sin \omega = \frac{\delta z /z }{(1/z +1) {\Delta} g(z) },
\end{equation}
{where $\delta z = {z}_\mathrm{obs}-z$ and $\Delta g(z)=g(z)-g(0)$. Further, using Eq. \eqref{eq:zBarHubble} it can be shown that $\delta z/z =\delta H_0/H_0$, with $\delta H_0 = H^{\rm obs}_0-H_0$.} 

{From} \citep{Reid:2019ApJ,Planck:2020_results2018}, we  have ${H^{\rm obs}_0}=73.5 \hunit$ at $7.57$ Mpc  and $H_0 =67.36 \hunit$. This gives $\delta H_0/H_0 =0.0911$. {Following standard $\Lambda$CDM and cosmological parameters from Planck 2018 results \citep{Planck:2020_results2018}, the proper distance $d=7.57$ Mpc corresponds to $z=0.0017$.}
So we set $\delta z /z = 0.0911$ with $z=0.0017$ in Eq. \eqref{eq:asinw} and obtain $\alpha \sin \omega= 0.2697$. {This value corresponds to} the red solid line of Figure \ref{fig:H_Reid_etal}, satisfying Reid et. al. \HL parameter observation.  

In addition to the above condition, parameters $\alpha$ and $\omega$ are also constrained by CMB quadrupole  and octopole values \citep{Erickcek:2008a}, 
\begin{equation}
\label{eq:quad} 
|\alpha_{\rm dec} \sin \omega| \le 5.8 \mathcal{Q} / (\kappa \chi_{_{\rm dec}})^2, 
\end{equation}
\begin{equation}
\label{eq:octo}
|\alpha_{\rm dec} \cos \omega| \le 32 \mathcal{O} / (\kappa \chi_{_{\rm dec}})^3, 
\end{equation}
where subscript `dec' denotes the parameters at decoupling, $\Psi_{\rm dec}=0.937 \Psi_p$ \citep{Erickcek:2008a} and thus $\alpha_{\rm dec}= 0.937 \alpha$.  $\chi_{_{\rm dec}}$ is the comoving distance to decoupling,  $\mathcal{Q}$ and $\mathcal{O}$ are 3 times the measured rms values of the CMB quadrupole and octopole, respectively \citep{Erickcek:2008a}. We use the latest Planck 2018 \citep{Planck:2020_results2018} values of $Q = 3\sqrt{C_2} \le 1.69 \times 10^{-5}$ and $\mathcal{O}=3\sqrt{C_3}\le 2.44 \times 10^{-5}$. Following Eq. \eqref{eq:quad}, we obtain the CMB limit on $|\alpha \sin \omega|$, i.e., the blue-shaded region of Figure \ref{fig:H_Reid_etal}. To explain Reid et al. \citep{Reid:2019ApJ} excess $H_0$, we  
 impose (a) $\alpha \sin \omega =0.2697$ and (b) CMB limits. Using these, we find  $\kappa/H_0 \le 6.29 \times 10^{-3}$. 

The NVSS galaxies extend over the redshift range $z=0$ to 3.5 \citep{Adi:2015nb,Tiwari:2016adi}. We note from Das et. al. \citep{Das:2021JCAP} that to 
obtain NVSS dipole as a consequence of superhorizon mode, 
\begin{equation}
\label{eq:acos_d}
\alpha \cos \omega = \frac{H_0}{\kappa} \frac{\mathcal{D}_{\rm obs} - \mathcal{B}}{\mathcal{A}_1(0,3.5) +\mathcal{A}_2(0,3.5)+\mathcal{C}(0,3.5)}, 
\end{equation}
where $\mathcal{D}_{\rm obs}$ is the observed NVSS dipole. 
Other terms in Eq. \eqref{eq:acos_d} are defined in Das et al. \citep{Das:2021JCAP}. To obtain $\alpha \sin \omega$ versus $\kappa/H_0$ curves (dashed-dotted curves in Figure \ref{fig:H_Reid_etal}) satisfying NVSS dipole, we multiply Eq. \eqref{eq:acos_d} by $\tan \omega$ after choosing a specific phase $\omega$. For the allowed $\kappa$ range, 
the value of $\alpha \cos \omega$ needed to explain excess NVSS dipole in Eq. \eqref{eq:acos_d} is much less than the octopole limit in  Eq. \eqref{eq:octo}.

The intersection points of red solid line and NVSS curves in Figure \ref{fig:H_Reid_etal} are the possible solutions to explain both \HL tension and NVSS excess dipole. The parameters for these superhorizon modes are listed in Table \ref{tab:SH_par}. 

In Figure \ref{fig:H_dist}, we fix $\alpha \sin \omega$ to satisfy Reid et. al. \citep{Reid:2019ApJ} \HL parameter value and then calculate {$H^{\rm obs}_0$} {predicted} with superhorizon mode from data at different distances (redshifts) using Eq. \eqref{eq:asinw}.

\renewcommand*{\bibfont}{\footnotesize}
\bibliographystyle{JHEP}
\bibliography{master_Hubble,master_radio}
\end{document}